\documentclass[11pt]{article}
\usepackage{graphicx}
\begin{document}
\bibliographystyle{unsrt}
\def\ra{\rangle}
\def\la{\langle}
\def\aao{\hat{a}}
\def\aaot{\hat{a}^2}
\def\aco{\hat{a}^\dagger}
\def\acot{\hat{a}^{\dagger 2}}
\def\ano{\aco\aao}
\def\bao{\hat{b}}
\def\baot{\hat{b}^2}
\def\bco{\hat{b}^\dagger}
\def\bcot{\hat{b}^{\dagger 2}}
\def\bno{\bco\bao}
\def\beqn{\begin{equation}}
\def\eeqn{\end{equation}}
\def\bear{\begin{eqnarray}}
\def\eear{\end{eqnarray}}
\def\cdott{\cdot\cdot\cdot}
\def\bcen{\begin{center}}
\def\ecen{\end{center}}
\def\nbar{\bar{n}}
\def\eps{\epsilon}
\def\hrho{\hat{\rho}}
\def\rhom{\hat{\rho}_m}
\def\rhot{\hat{\rho}_t}
\def\rhod{\hat{\rho}_d}
\title{Effect of thermal noise on atom-field interaction: Glauber-Lachs versus Mixing}
\author{
S. Sivakumar\\Materials Physics Division\\ 
Indira Gandhi Centre for Atomic Research\\ Kalpakkam 603 102 INDIA\\
Email: siva@igcar.gov.in\\
Phone: 91-044-27480500-(Extension)22503}
\maketitle
\begin{abstract}
Coherent signal containing thermal noise is a mixed state of radiation.  There are two distinct classes of such states, a Gaussian state obtained by Glauber-Lachs mixing and a non-Gaussian state obtained by the canonical probabilistic mixing of thermal state and coherent state.   Though both these versions are noise-included signal states, the effect of noise is less pronounced in the Glauber-Lachs version.  Effects of these two distinct ways of noise addition is considered in the context of atom-field interaction;  in particular, temporal evolution of population inversion and atom-field entanglement are studied.  Quantum features like the collapse-revivals in the dynamics of population inversion and entanglement are diminished by the presence of thermal noise.   It is shown that the features lost due to the presence of thermal noise are restored by the process of photon-addition.
\end{abstract}
\newpage
\section{Introduction}\label{secI}
Thermal radiation is a source of noise in the  context of coherent signal.  This source of noise is unavoidable at finite temperatures, making it necessary to understand its effects on experiments.  
Thermal radiation is characterized by the temperature ($T$) of the source emitting the thermal photons.   The density operator for a single mode field in thermal state\cite{gerryknight} is  
\beqn
\hat{\rho}_t=\frac{1}{1+\nbar}\sum_{n=0}^\infty\left[\frac{\nbar}{1+\nbar}\right]^n\vert n\ra\la n\vert.
\eeqn 
The average number of thermal photons is $\nbar$, which is related to the temperature of the source of thermal photons.  The photon-number distribution $\la n\vert\hat{\rho}_t\vert n\ra$ is a monotonically decreasing function of $n$, that is, the maximum probability is for the vacuum state.  On the other hand, 
a coherent state of radiation field is a pure state which is characterized by a complex number $\alpha$. In the number state basis, the coherent state of amplitude $\alpha$\cite{gerryknight} is 
\beqn
\vert\alpha\ra=\exp\left[-\frac{\vert\alpha\vert^2}{2}\right]\sum_{n=0}^\infty\frac{\alpha^n}{\sqrt{n!}}\vert n\ra.
\eeqn
Unlike the thermal state, a coherent state has a single peaked photon-number distribution, and the  peak of the distribution occurs for $n\approx\vert\alpha\vert^2$.  Since the thermal state is a mixed state, except at $T=0$K, any state of light that incorporates thermal noise is also a mixed state. Thermal radiation can be incorporated into the coherent radiation {\it via} either  
 mixing or Glauber-Lachs (GL) superposition of coherent state and thermal field.  The GL 
version corresponds to the unitarily displaced thermal state (DTS)\cite{Glauber, Lachs},  
\beqn\label{GLS}
\hrho_{_d}=\hat{D}(\alpha)\hat{\rho}_{_t}\hat{D}^\dagger(\alpha)
\eeqn
where $\hat{D}(\alpha)=\exp[\alpha\aco-\alpha^*\aao]$ is the displacement operator expressed in terms of the creation operator $\aco$ and the annihilation operator $\aao$ of the field.   These states are the quantum versions of  classical channels with additive Gaussian noise\cite{cavesdrummond}.  The DTS can be charaterized as an intermediate state between the thermal state and the coherent state, the two limiting cases of the DTS corresponding to $\nbar=0$ and $\alpha=0$ respectively \cite{valverde}.  
The other scheme to incorporate thermal noise is mixing, which is relevant if the field is obtained by probabilistically choosing 
 photons from a coherent source and a thermal source.   The resultant state is described by 
\beqn\label{mtcs}
\hrho_m=(1-q)\hrho_t+q\vert\alpha\ra\la\alpha\vert,
\eeqn
where $q$ is  non-negative and less than unity.  Such  states are relevant in the study of noisy quantum channels that transmit either the coherent state $\vert\alpha\ra$ with probability $q$ or the thermal state $\rho_{t}$ with probability $1-q$.  In the subsequent discussion,  the state $\hrho_m$ is referred as mixed thermal-coherent states (MTCS).

\begin{figure}
\centering
\includegraphics[height=8cm,width=10cm]{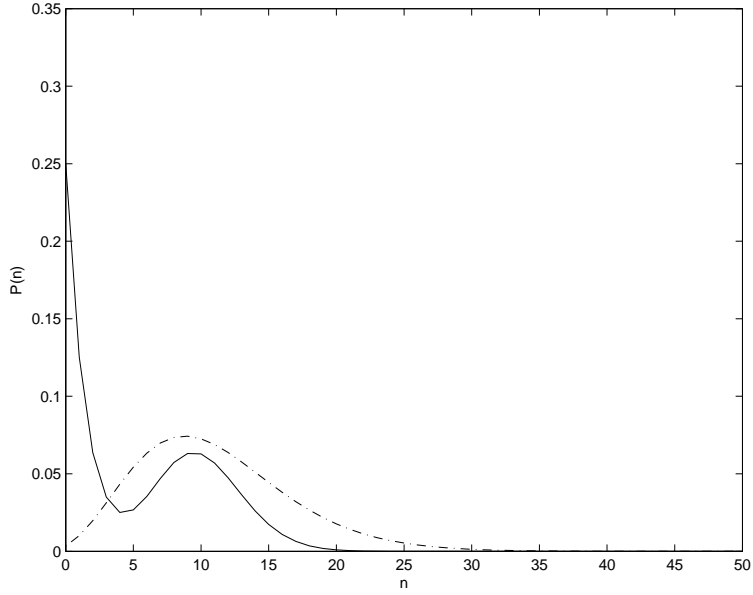}
\caption {Photon-number distribution $P(n)$ for DTS  (dashed) and MTCS (continuous).  The amplitude $\alpha$ is $\sqrt{10}$ and $q=0.5$.}
\label{fig:Fig1}
\end{figure}

The DTS and the MTCS, albeit constructed out of the coherent states and the thermal states,  are not equivalent.  The DTS is a Gaussian state while the MTCS, which is a mixture of Gaussian states, namely, the coherent state and the thermal state,  is not a Gaussian state unless $q=0,1$\cite{paris}.   
        If the mean photon-number $\nbar$ is zero, the DTS becomes the coherent state $\vert\alpha\ra$, a pure state with no thermal noise; in the same limit, the MTCS is a mixture of $\vert\alpha\ra$ and the vacuum state $\vert 0\ra$.  It is further required to have $q=1$ to ensure that there is no thermal noise in the MTCS.    If the amplitude $\alpha$ vanishes, the DTS becomes the thermal state $\rho_{t}$ while the MTCS is a mixture of thermal state and vacuum state.     In Fig.\ref{fig:Fig1}, photon-number distribution is shown for DTS (dashed curve) and MTCS (continuous curve) corresponding to $\alpha=\sqrt{10}$ and $q=0.5$.  The photon number distribution of DTS exhibits a single peak, whereas MTCS has one at $n=0$ and the location of the other peak depends on the values of $\alpha$ and $\nbar$. 

The paper is organized as follows: in Section II, the DTS is shown to be a mixture of photon-added coherent states.  In Section III, interaction of a two-level atom with the radiation field is studied.  The objective of the study is to highlight the differences in the evolution of entanglement and population inversion when the atom interacts with the DTS and the MTCS.  In Section IV, effects of  photon-addition on atom-field dynamics are discussed, followed by a summary of the results.

\section{Glauber-Lachs states as mixture of photon-added coherent states}

      A photon-added coherent state (PACS) of order $m$, is defined as\cite{gsatara}
\beqn
\vert\alpha,m\ra=\frac{\aao^{\dagger m}\vert\alpha\ra}{\sqrt{m!L_m(-\vert\alpha\vert^2)}}.
\eeqn  
Here $L_m(x)$ is the $m$th order Laguerre polynomial\cite{grad} and $m$ is a nonnegative integer.  The first index $\alpha$ refers to the complex amplitude and the second index $m$ is the order of the PACS.  
 It is important to note that PACS of different orders are not orthogonal to each other.  These states have been studied, both theoretically and experimentally, in a variety of contexts, such as  nonclassicality and quantum-classical correspondence\cite{zavatta}.   In this section it is shown that DTS is expressible as a mixture of all orders of PACS, all of same amplitude.

    The definition of DTS given in Eq. \ref{GLS} yields the following expression
\beqn\label{mdncs}
\hrho_{d}=\frac{1}{1+\nbar}\sum_{n=0}^\infty\left[\frac{\nbar}{1+\nbar}\right]^n\vert n,\alpha\ra\la n,\alpha\vert,
\eeqn  
which is a mixture of orthogonal states, namely, the displaced number states $\vert n,\alpha\ra=\hat{D}(\alpha)\vert n\ra$.  Another 
way of expressing the displaced thermal states is to note that 
\beqn
\hrho_d=\frac{1}{1+\nbar}\left[\frac{\nbar}{1+\nbar}\right]^{(\aco-\alpha^*)(\aao-\alpha)}.
\eeqn
Defining $\nbar=1/(\exp^\gamma-1)$, the above expression for $\hrho_d$ can be simplified to \cite{loui}
\beqn
\hrho_d=\eps e^{-\eps\vert\alpha\vert^2} e^{\eps\alpha\aco} e^{-\gamma\aco\aao} e^{\eps\alpha\aao}.
\eeqn
The parameter $\eps$ is given by $1/(1+\nbar)$.  Expanding $e^{-\gamma\aco\aao}$ in terms of the Fock state 
projectors $\vert n\ra\la n\vert$, the density operator of the DTS becomes,
\beqn
\hrho_{d}=\eps e^{-\eps\vert\alpha\vert^2}\sum_{n=0}^\infty \frac{e^{-\gamma n}}{n!}e^{\eps\alpha\aco}\hat{a}^{\dagger n}\vert 0\ra\la 0\vert\aao^n e^{\eps\alpha^*\aao}.
\eeqn
Since $\exp(\alpha\aco)$ [respectively, $\exp(-\alpha^*\aao)$] and $\aco$ [respectively, $\aao$] commute, the order of multiplication can be reversed. It is easy to recognize that $\exp(-\epsilon\alpha\aco)\vert 0\ra= \exp(\eps^2\vert\alpha\vert^2/2)\vert\eps\alpha\ra$, a coherent state of amplitude $\eps\alpha$, apart from  the multiplication factor $\exp(\eps^2\vert\alpha\vert^2/2)$.  
   Further,  $\hat{a}^{\dagger n}\vert\eps\alpha\ra$ are PACS, apart from the normalization factor $\sqrt{n!L_n(-\eps\vert\alpha\vert^2)}$.  With these identifications, the previous equation is recast as  
\beqn\label{mpacs}
\hrho_d=\frac{e^{-\left|\tilde\alpha\right|^2\frac{\nbar}{1+\nbar}}}{1+\nbar} \sum_{n=0}^\infty \left[\frac{\nbar}{1+\nbar}\right]^n L_n(-\left|\tilde\alpha\right|^2)\left|{\tilde\alpha,n}\rangle\langle\tilde\alpha,n\right|,
\eeqn
where $\tilde\alpha=\alpha/(1+\nbar)$.
The expression reveals that DTS  is a mixture of all orders of PACS of amplitude $\alpha/(1+\nbar)$.  It is interesting to contrast the two expressions for the DTS, given in Eqns. \ref{mdncs} and \ref{mpacs} respectively: the former relation expresses the DTS as a mixture of orthogonal states while the later expresses it as a mixture of nonorthogonal states.

\section{Population inversion and entanglement dynamics}

In this section, dynamics of a two-level atom interacting with a single mode field is considered.  The initial state of the field is either DTS or MTCS.  The presence of thermal noise in the initial state  affects the dynamics of interaction.  Evolution of atom-field state is studied to understand the effects of different ways of incorporating the thermal noise.  The atom-field interaction is modelled by the  Jaynes-Cummings hamiltonian\cite{jaynes} 
\beqn
\hat{H}_{JC}=\hbar\omega\aco\aao+\frac{\hbar\nu}{2}\sigma_z+\hbar\lambda(\aco\sigma_-+\aao\sigma_+),
\eeqn
where $\nu$ is the atomic transition frequency, $\omega$ is the field frequency and $\lambda$ is the coupling constant for the atom-field interaction.  
Using a suitable interaction picture and rotating wave approximation\cite{Bradmore}, the hamiltonian is 
\beqn
\hat{H}_I=\hbar\frac{\Delta}{2}\sigma_z+i\hbar\lambda\left(\aco\sigma_--\aao\sigma_+\right),
\eeqn
where $\Delta=\omega-\nu$ is the detuning parameter.  The operators relevant to the two-level atom are the Pauli operator $\sigma_z$ and, the raising and lowering operators $\sigma_-$ and $\sigma_+$ respectively.

The  time-evolved density operator $\hat{\rho}(t)$ is $\hat\rho(t)=\hat{U}(t)\hat\rho(0)\hat{U}^\dagger(t)$, where $\hat{\rho}(0)$ is the initial density operator of the atom-field system. The unitary evolution operator $\hat{U}(t)$ is $\exp\left[-i\hat{H}_I t/\hbar\right]$.    In this report, the initial state of the atom-field system is taken to be $\vert e\ra\la e\vert\otimes\rho_f$, a factorizable state which has no entanglement.   The operator $\rho_f$ stands for the field state which is taken to be either DTS or MTCS.   The states $\vert e,n\ra$ and $\vert g,n+1\ra$ transform under the action of $\hat{H_I}$ as follows:
\bear
\hat{H_I}\vert e,n\ra&=&\hbar\frac{\Delta}{2}\vert e,n\ra+i\hbar\lambda\sqrt{n+1}\vert g,n+1\ra,\\
\hat{H_I}\vert g,n+1\ra&=&\hbar\frac{\Delta}{2}\vert g,n+1\ra-i\hbar\lambda\sqrt{n}\vert e,n\ra.
\eear
The results  indicate that the span of the states $\vert e,n\ra$ and  $\vert g,n+1\ra$, is invariant under the action of  the unitary evolution operator $\exp(-it\hat{H_I}/\hbar)$.   If the interaction is resonant, ($\Delta=0$), the time-evolved density operator  is 
\beqn
\hrho(t)=\sum_{n,m=0}^\infty\rho_{n,m},
\eeqn
with
\beqn
\rho_{n,m}=c_{nm}\left(
\begin{array}{c}
\cos(\lambda\sqrt{n}t)\cos(\lambda\sqrt{m}t)\\
\sin(\lambda\sqrt{n}t)\cos(\lambda\sqrt{m}t)\\
\cos(\lambda\sqrt{n}t)\sin(\lambda\sqrt{m}t)\\
\sin(\lambda\sqrt{n}t)\sin(\lambda\sqrt{m}t)\\
\end{array}
\right)^{\tau}
\left(
\begin{array}{c}
\vert e,n\ra\la e, m\vert\\
\vert g,n+1\ra\la e, m\vert\\
\vert e,n\ra\la g, n+1\vert\\
\vert g,n+1\ra\la g, n+1\vert
\end{array}
\right).
\eeqn
The superscript $\tau$ implies matrix transpose.  
The coefficients $c_{nm}$ are the matrix elements $\la n\vert\hrho\vert m\ra$ of the initial density operator of the field; the relevant expressions for the DTS and the MTCS are
\bear
\la n\vert\hat{\rho}_m\vert m\ra&=&\frac{q}{1+\nbar}\left(\frac{\nbar}{1+\nbar}\right)^n\delta_{n,m}+(1-q)\exp(-\vert\alpha\vert^2)\times\nonumber\\
& &~~\exp[i(n-m)\theta]\frac{\vert\alpha\vert^{n+m}}{\sqrt{n!m!}},\\
\la n\vert\hat{\rho}_d\vert m\ra &=&\sqrt{\frac{n!}{m!}}\left[\frac{1}{1+\nbar}\right]^{m-n+1}\left[\frac{\nbar}{1+\nbar}\right]^n\exp[-\vert\alpha\vert^2/(1+\nbar)]\times\nonumber\\
&&~~\exp[i(n-m)\theta]\vert\alpha\vert^{m-n}L_m^{m-n}\left(\frac{\vert\alpha\vert^2}{\nbar(1+\nbar}\right),
\eear 
respectively. It is also clear that $c_{mn}=c_{nm}^*$, the complex conjugate of $c_{nm}$.

For the purpose of comparing the interaction dynamics of the atom with the two classes of the  field, namely, the DTS and the MTCS, it is essential that a parameter is chosen to compare the two fields.   While the displacement $\alpha$ and mean number of thermal photons $\nbar$ are common to both the states, the parameter $q$ in the MTCS is not fixed.  A meaningful way is to choose the value $q$ so that the overlap with the coherent state $\vert\alpha\ra$ is the same for both the DTS and the MTCS, {\em i.e.}, $\la\alpha\vert\rho_m\vert\alpha\ra=\la\alpha\vert\hat{\rho}_t\vert\alpha\ra$.  With this restriction,  the value of $q$ is fixed by the parameters $\alpha$ and $\nbar$.  The respective contributions of coherent state $\vert\alpha\ra$  to the DTS and MCTS are 
\bear
\la\alpha\vert\hat{\rho}_m\vert\alpha\ra&=&q+\frac{1-q}{1+\nbar}\exp[-\frac{\vert\alpha\vert^2}{1+\nbar}],\\
\la\alpha\vert\hat{\rho}_d\vert\alpha\ra&=&\frac{1}{1+\nbar}.
\eear 
The contribution from the coherent state $\vert\alpha\ra$ to the states  decreases as $\nbar$ increases.  Let $\bar{q}$ that special value of $q$ which ensures equal overlap of the DTS and MTCS with the coherent state $\vert\alpha\ra$.  Then, the last two expressions yield
\beqn
\bar{q}=\frac{1-\exp[-\vert\alpha\vert^2/(1+\nbar)]}{\nbar+1-\exp[-\vert\alpha\vert^2/(1+\nbar)]}.
\eeqn
In physical terms, this is the probability with which coherent state photon  is chosen in the mixed state  defined in Eq.\ref{mtcs} to ensure equal overlap with $\vert\alpha\ra$.  If $\nbar$ is zero (no thermal photons), $\bar{q}$ becomes  unity.  Hence, noise-free limit of the MTCS is the pure coherent state $\vert\alpha\ra$, as in the case of DTS.  If $\alpha=0$, then $\bar{q}$ is $0$.  In this limit, the states $\rhod$ and $\rhom$ become the thermal state $\rhot$, a mixed state.  Thus the condition of equal overlap implies that the limiting cases of the DTS and MTCS are the same.  This implies that the  DTS and the MTCS are two classes of states that interpolate between a pure state ($\vert\alpha
\ra$)and a mixed state (thermal state). \\

During evolution, the state of the atom and the field is entangled,  a feature that originates in the  bipartite nature of the system.    The entanglement between the atom and the field is quantified by the logarithmic negativity $N$, defined as the absolute sum of the negative eigenvalues of the partially transposed density operator $\hat{\rho}^{PT}$\cite{wei}.  If $\lambda_k$ are the eigenvalues of $\hat{\rho}^{PT}(t)$, then $N(t)=\sum_k\left[\vert\lambda_k\vert-\lambda_k\right]/2$.  In the case of atom-field evolution under Jaynes-Cummings interaction, the time-averaged entanglement depends on the initial mixedness of the bipartite state\cite{Kayhan}.  The mixedness of the DTS is 
\beqn
1-Tr[\hrho_d^2]=\frac{2\nbar}{1+2\nbar}.
\eeqn
This expression is independent of $\alpha$ and hence the mixedness of the DTS is same as that of the thermal state.  This is a consequence of the fact that the two states are related by a unitary transformation.  The mixedness of the MTCS, given by
\beqn
1-Tr[\hrho_m^2]=1-q^2-\frac{(1-q)^2}{1+2\nbar}-\frac{2q(1-q)}{1+\nbar}\exp[-\frac{\vert\alpha\vert^2}{1+\nbar}],
\eeqn
depends on $\nbar$ and $\alpha$.

As far as the atomic dynamics is concerned, the relative probabilities  to be in the excited and states are important.  In this context, A quantitative  population inversion $W(t)$, defined as the difference of the probabilities for the  atom to be in the excited state and 
ground state, is of interest.  Since the atomic Hamiltonian is proportional to $\vert e\ra\la e\vert-\vert g\ra\la g\vert$, the 
population inversion is essentially the energy of the atom, apart from an  overall scaling  factor $\hbar\nu$.  
If the state of the field is a coherent state of suitably large amplitude, the time-evolution of $W(t)$ is characterized by the well known "collapse-revival" structures\cite{cummings,eberly, walther}.  There are  associated features present in the evolution of entanglement between the atom and the field\cite{gerryknight}.  The effectiveness of the thermal radiation in diminishing these features of atom-field is a qualitative indicator of the effects of thermal noise \cite{mvs}.

If $\nbar$ and $\vert\alpha\vert$ are less than unity, both the DTS and the MTCS are approxiamtely the vacuum state of the field.  The interaction dynamics of a two-level atom in its excited state with the vacuum state leads to a periodic exchange of energy between the field and the atom.   Therefore, nontrivial dynamics can be expected if $\vert\alpha\vert$ differs significantly from unity.  As a representative case, the numerical values considered for the study in this work are: $\vert\alpha\vert^2=10$ and $\nbar$ is taken to be a fraction of $\vert\alpha\vert^2$.

\begin{figure}
\centering
\includegraphics[height=8cm,width=10cm]{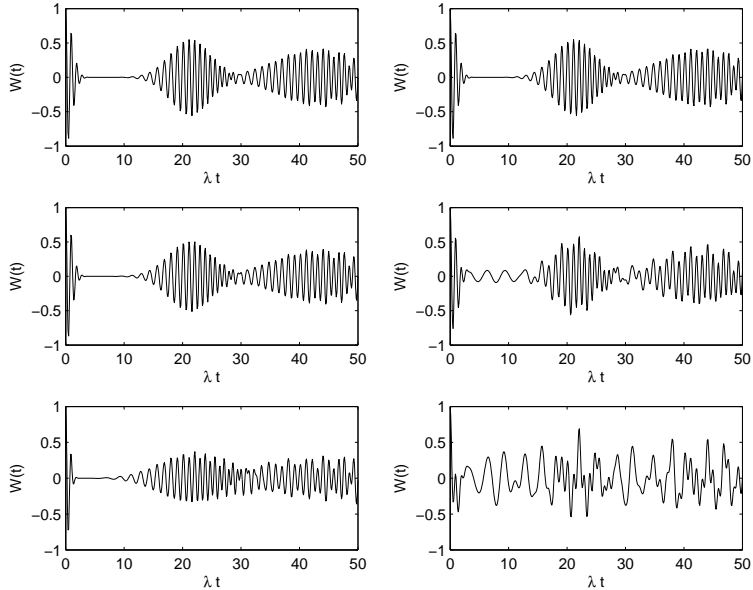}
\caption {Time evolution of population inversion $W(t)$.  In all the subplots, the abscissa corresponds to the scaled time $\lambda t$. Left column corresponds to evolution with DTS as the initial state of the field  and the right column corresponds to MTCS as the initial state.  For the both the states $\alpha=\sqrt{10}$.  Different rows correspond to different values of mean number of thermal photons: $\nbar=0$ (first row), 0.1(second) and 1.0 (third). }
\label{fig:Fig2}
\end{figure}

\begin{figure}
\centering
\includegraphics[height=8cm,width=10cm]{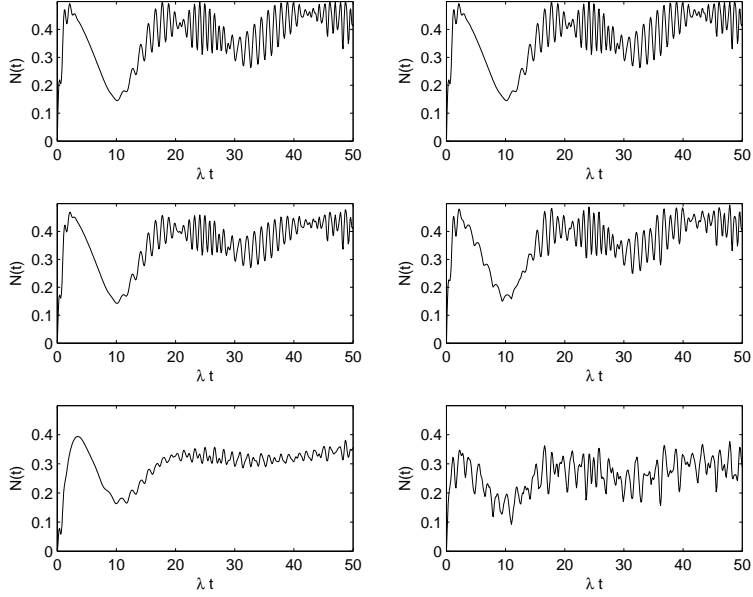}
\caption {Time evolution of negativity $N(t)$.    Other descriptions are as given in Fig. \ref{fig:Fig1}.  
}
\label{fig:Fig3}
\end{figure}

In Fig. \ref{fig:Fig2}, the temporal variation of population inversion is shown for three different values of $\nbar$.   For easy comparison, the corresponding profiles for the DTS and MTCS are shown in adjacent columns.  
In the noise-free limit, that is, $\nbar=0$, the DTS and the MTCS are identical and there is no difference in the  dynamics of interaction with the two-level atom as seen from the first row of figures.  However, as $\nbar$ is increased to 0.1, equal to one percent of the coherent state photon number, the dynamics in the DTS case is very different from that of the MTCS.  This is seen by comparing the two figures in  second row. 
The profiles clearly show the marked differences in the dynamics induced by the two different initial states.  As the fraction of thermal photons is increased to ten percent ($\nbar=1$), the temporal profile of the population inversion in the case of  MTCS exhibits higher degree of deviation from the $\nbar=0$ case, notably, the collapse-revival structure is totally absent.   In contrast, in the case of DTS the collapse-revival structure is still intact even if $\nbar=1$.    This indicates that the thermal noise in the MTCS is more effective than that in the Glauber-Lachs version.      
The temporal evolution of negativity $N(t)$ is shown in Fig. \ref{fig:Fig3}.  To compute the eigenvalues of the partially transposed bipartite density matrix to high accuracy, the size of the density matrix has been chosen to be 300X300.  Since the initial photon-distribution is appreciable for $n\approx10$ and falls rapidly for larger values of $n$, the chosen size is large enough to ensure that the computed eigenvalues are not spurious.  The first, second and third rows of figures in Fig. \ref{fig:Fig3} correspond to $\bar{n}=0.1,1$ and $n=1$ respectively.  As in the case of population inversion,  the dynamics of $N(t)$ shows significant changes in the case of MTCS as the value of $\nbar$ increases, leading to complete loss of structures in the temporal profile.  The Glauber-Lachs mixing too leads to changes in the evolution of $N(t)$ as noise level increases, though to a lesser extent than in the MTCS case.   Thus, atom-field entanglement is more robust against thermal noise if the thermal noise is included through Glauber-Lachs version rather than mixing.  It may be noted that if $\vert\alpha\vert$ is large, then the condition of equal overlap with coherent state also implies that the states are of equal mixedness.  

\section{Noise reduction by photon-addition}

      In the previous section the effect of thermal noise on the dynamics of atom-field interaction was discussed.  In this section, effects of photon-addition are presented.  One of the consequences of photon-addition is that the resultant state has no overlap with the vacuum state $\vert 0\ra$.  In the thermal state of the radiation field,  the vacuum state probability is not less than that of other number states.  Since the photon-added states do not have  any contribution from the vacuum state, it can be expected that the effects of noise are less pronounced.  
 
       The photon-added versions of the DTS and MTCS are 
\bear
\tilde\rho_m&=&\frac{\aco\rho_m\aao}{Tr[\aco\rho_m\aao]},\\
\tilde\rho_d&=&\frac{\aco\rho_d\aao}{Tr[\aco\rho_d\aao]},
\eear
wherein the photon-added states are indicated with a tilde.

Photon-addition introduces nonclassical features such as sub-Poissonion statistics, squeezing, etc, apart from increasing the mean number of photons\cite{ctlee, kiesel,arusha}.  Additionally,  effect of   photon-addition shows up in the atom-field interaction dynamics too.  In particular, the features lost due to the presence of thermal photons are restored as a consequence of photon-addition.  The overlap between the  coherent state $\vert\alpha\ra$ and the 
photon-added versions of the DTS and MTCS are related to the corresponding overlaps of the DTS and MTCS {\it via}, 
\bear
\la\alpha\vert\tilde\rho_m\vert\alpha\ra&=&\frac{\vert\alpha\vert^2}{1+\bar{q}\nbar+(1-\bar{q})\vert\alpha\vert^2}\la\alpha\vert\hat{\rho}_m\vert\alpha\ra,\\
\la\alpha\vert\tilde\rho_d\vert\alpha\ra&=&\frac{\vert\alpha\vert^2}{1+\nbar+N_c}\la\alpha\vert\hat{\rho}_d\vert\alpha\ra.
\eear
If $\vert\alpha\vert^2=10$ and $\nbar=1$,  the probability $\bar{q}$ is nearly 0.5.  It is seen  that the overlap of the photon-added MTCS $\tilde\rho_m$ with the coherent state is significantly larger compared to $\la\alpha\vert\rho_m\vert\alpha\ra$.  So, it is natural to expect the dynamics of atom-field interaction to be similar to that in the case of field being in a coherent state.  It is interesting to note that in the case of $\tilde\rho_t$, photon-addition leads to a  decrease in the overlap with coherent state.   

\begin{figure}
\centering
\includegraphics[height=8cm,width=10cm]{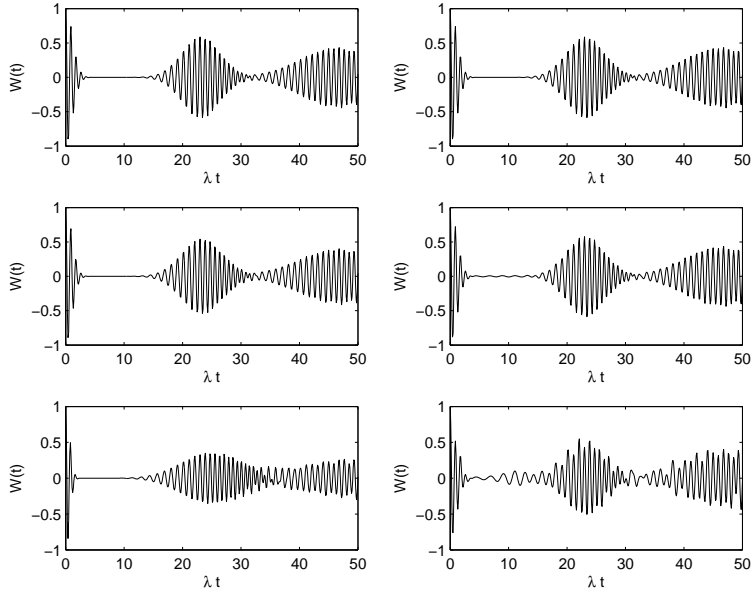}
\caption {Time evolution of population inversion in the interaction between atom and photon-added states. The abscissa corresponds to the scaled time $\lambda t$.  Left column corresponds to evolution with the photon-added DTS $\tilde\rho_d$ as the initial state of the field and the right column corresponds to $\tilde\rho_m$ as the initial state. For the both the states $\alpha=\sqrt{10}$.  Different rows correspond to different values of mean number of thermal photons: $\nbar=0$ (first row), 0.1(second) and 1.0 (third). }
\label{fig:Fig4}
\end{figure}

The evolution  of population inversion in the case of photon-added states shown in Fig.\ref {fig:Fig4}.  The effects of photon-addition on the dynamics of $W(t)$ is clear on comparing the corresponding profiles  in Fig. \ref{fig:Fig4} and  Fig. \ref{fig:Fig2}.  Noise does not modify the profiles much if the initial state is the DTS or its photon-added version. However, comparing the right columns of the two figures, it is clear that for the MTCS, the effect of photon-addition is very significant.  In particular, with $\nbar=1$, the evolution of $W(t)$ in the  photon-added version is very similar to the zero noise, that is, $\nbar=0$ case.  

\begin{figure}
\centering
\includegraphics[height=8cm,width=10cm]{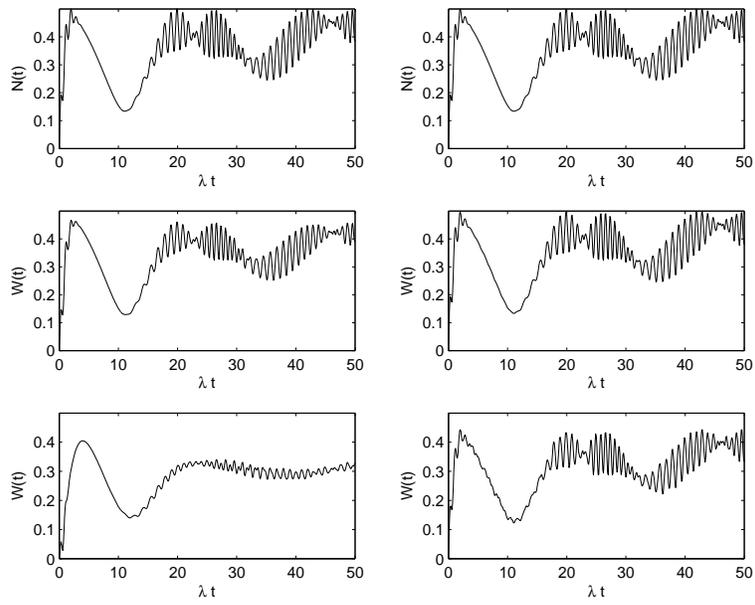}
\caption {Time evolution of negativity $N(t)$ due to  the interaction between atom and  photon-added states. Other descriptions are same as given in Fig. \ref{fig:Fig4}}
\label{fig:Fig5}
\end{figure}

	In Fig. \ref{fig:Fig5}, evolution profiles of  negativity if the initial state is one of the photon-added states  are shown.   To assess the effect of photon-addition, the corresponding  profiles in Figs \ref{fig:Fig3} and \ref{fig:Fig5} are compared.  Specifically, the right columns of the two figures show the recourse to noise-free evolution due to photon-addition to the MTCS.  As in the case of $W(t)$, the effect of photon-addition is very mild in the case of DTS.  Though the process of photon-addition is applied to both the coherent component and the thermal component of the MTCS, the net effect is to produce effects similar to that of coherent radiation as far as the atom-field dynamics is concerned.  In short, the noise effect is suppressed to a large extent resulting in the recovery of the collapse-revival structures present in the  noise-free evolution.  

The reason for the recovery of noise-free evolution on photon-addition is due to the the fact that the vacuum state probability is vanishes.  The photon-added states $\tilde\rho_m$ and $\tilde\rho_d$ have no overlap with the vacuum state $\vert 0\ra$, that is, $\la 0\vert\tilde\rho\vert 0\ra=0$, where $\tilde\rho$ refers to the photon-added versions of the DTS and  MTCS.    When the mean number of thermal photons is about unity, the major contribution to the thermal state is from the vacuum.  In the photon-added versions, the contribution from the vacuum is absent, resulting in lesser overlap with the thermal state.  This, in turn, means that the overlap with coherent states is more for the photon-added versions of the MTCS and DTS.  

\section{Summary}
Thermal states and coherent states are Gaussian states.  While displacing the thermal state is a way of incorporating thermal noise, and the resultant state is still Gaussian.  Mixing of thermal and coherent states is another way of including thermal noise, however, the resultant state is non-Gaussian.  The displaced thermal state is a mixture of photon-added coherent states of all orders.  Though the two classes of states are closely related to thermal and coherent states, it is only for a special choice of mixing probability in the mixture of thermal and coherent states,  these two classes of states have equal overlap with coherent states.   In the case of displaced thermal state, the dynamics of population inversion and atom-field entanglement are not very susceptible to thermal photons if the coherent state amplitude is large $(\alpha\approx\sqrt{10})$ and the thermal photons contribution is as high as one tenth of the total number of photons.  In the case of MTCS, though it has the same overlap with the coherent state $\vert\alpha\ra$ as with the case of DTS, increasing the thermal photon contribution renders the dynamics totally devoid of any collapse-revival structure in the evolution of population inversion and atom-field entanglement.  Photon-added versions of these two classes of states exhibit, in the context of atom-field interaction,  features very similar to the case of dynamics in the absence of thermal noise.  In essence, photon-addition suppresses the effects of thermal noise, thereby amplifying the effect of coherent states on the dynamics.


\begin{thebibliography}{40}
\bibitem{gerryknight}{C. C. Gerry and P. L. Kinght, Introductory Quantum Optics, Cambridge University Press, New York (2005). }
\bibitem{Glauber}{R. J. Glauber in Physics of Quantum Electronics; Conference proceedings. Edited by P. L. Kelley, B. Lax and P. E. Tannenwald, McGraw Hill (1966).}
\bibitem{Lachs}{G. Lachs, Phys. Rev. {\bf 138 B} (1965)1012-1016.}
\bibitem{cavesdrummond}{C. M. Caves and P. B. Drummond, Rev. Mod. Phys. {\bf 66} (1994)481-538.}
\bibitem{valverde}{C. Valverde and B. Baseia, Int. J. Quant. Inf. {\bf 2} (2004) 421-445.}
\bibitem{paris}{M. G. Genoni and M. G. A. Paris, Phy. Rev. A {\bf 82} (2010) 052341 (1-19).}
\bibitem{gsatara}{G. S. Agarwal and K. Tara, Phys. Rev. A {\bf 43} (1991) 
492-497. }
\bibitem{grad}{I. S. Gradshteyn and I. M. Ryzhik, Table of Integral, Series and 
Products, Academic Press, USA, 2000.}
\bibitem{zavatta}{A. Zavatta, S. Viciani and M. Bellini, Science {\bf 306} 
(2004) 660-662.}
\bibitem{loui}{H. Louissell, Quantum Statsitical Properties of Radiation, John-Wiley, New York (1973).}
\bibitem{jaynes}{ E.T. Jaynes, F.W. Cummings, Proc. IEEE \bf{51} (1963) 89–109. }

\bibitem{Bradmore}{Barnett and Radmore, Methods of Theoretical Quantum Optics, Springer, NY, 1997.}
\bibitem{wei}{T. C. Wei et al, Phys. Rev. A {\bf 67} (2003) 022110(1-12).}
\bibitem{Kayhan}{H. Kayhan, Phys. Scr. {\bf 83} (2011) 025402(1-5).}

\bibitem{cummings}{F. W. Cummings, Phys. Rev. {\bf 140} (1965) A1051-A1056.}
\bibitem{eberly}{J. H. Eberly, N. B. Narozhny, and J. J. Sanchez-Mondragon, Phys. Rev. Lett. {\bf 44} (1980) 1323-1326.}
\bibitem{walther}{G. Rempe, H. Walther and N. Klein, Phys, Rev. Lett. {\bf 58} (1987)353-356. }
\bibitem{mvs}{M. V. Satyanarayana and M. Vijayakumar, Phys. Rev. A {\bf 45} (1992) 5301-5304.}
\bibitem{ctlee}{G. N. Jones, J. Haight and C. T. Lee, Quantum Semiclass. Opt. {\bf 9} (1997) 411-418.}
\bibitem{kiesel}{T. Kiesel, W. Vogel, M. Bellini and A Zavatta, Phys. Rev. A {\bf 83} (2011) 032116(1-5).}
\bibitem{arusha}{A. R. Usha Devi, R. Prabhu and M. S. Uma, Eur. Phys. J. D {\bf 40} (2006) 133-138.}
\end{thebibliography}
\end{document}